\documentclass[aps,prl,twocolumn,amsmath,amssymb,superscriptaddress,showpacs,10pt]{revtex4-1}
\newcommand{\etal}{\textit{et al}.}

\usepackage[caption=false]{subfig}
\usepackage{graphicx}
\usepackage{mathrsfs}
\usepackage{bm}
\usepackage[colorlinks]{hyperref}
\usepackage{braket}

\begin{document}

\newcommand{\ms}[1]{\mbox{\scriptsize #1}}
\newcommand{\msb}[1]{\mbox{\scriptsize $\mathbf{#1}$}}
\newcommand{\msi}[1]{\mbox{\scriptsize\textit{#1}}}
\newcommand{\nn}{\nonumber} 
\newcommand{\dg}{^\dagger}
\newcommand{\smallfrac}[2]{\mbox{$\frac{#1}{#2}$}}
\newcommand{\pfpx}[2]{\frac{\partial #1}{\partial #2}}
\newcommand{\dfdx}[2]{\frac{d #1}{d #2}}
\newcommand{\half}{\smallfrac{1}{2}}
\newcommand{\s}{{\mathcal S}}
\newcommand{\redtext}{\color{red}}
\newcommand{\bluetext}{\color{blue}}
\newcommand{\rws}{\color{blue}}
\newcommand{\kurt}{\color{green}} 
\newtheorem{theo}{Theorem} \newtheorem{lemma}{Lemma}

%
% Title
% 

\title{Fast, High-Fidelity, Quantum Non-demolition Readout of a Superconducting Qubit \\ Using a Transverse Coupling}

%
% Author
%

\author{Bryan T. Gard}
\email{bryantgard1@gmail.com}
\affiliation{U.S. Army Research Laboratory, Computational and Information Sciences Directorate, Adelphi, Maryland 20783, USA}
\affiliation{Hearne Institute for Theoretical Physics, Louisiana State University, Baton Rouge, LA 70803, USA}
\author{Kurt Jacobs}
\affiliation{U.S. Army Research Laboratory, Computational and Information Sciences Directorate, Adelphi, Maryland 20783, USA}
\affiliation{Hearne Institute for Theoretical Physics, Louisiana State University, Baton Rouge, LA 70803, USA}
\affiliation{Department of Physics, University of Massachusetts at Boston, Boston, MA 02125, USA}
\author{Jos\'{e} Aumentado}
\affiliation{National Institute of Standards and Technology, 325 Broadway St MS686.05, Boulder, Colorado 80305, USA}
\author{Raymond W. Simmonds}
\affiliation{National Institute of Standards and Technology, 325 Broadway St MS686.05, Boulder, Colorado 80305, USA}

%
% Abstract
%

\begin{abstract}
While relatively easy to engineer,  static transverse coupling between a qubit and a cavity mode satisfies the criteria for a quantum non-demolition (QND) measurement only if the coupling between the qubit and cavity is much less than their mutual detuning.  This can put significant limits on the speed of the measurement, requiring trade-offs in the circuit design between coupling, detuning, and decoherence introduced by the cavity mode. Here, we study a circuit in which the qubit-cavity and the cavity-feedline coupling can be turned on and off, which helps to isolate the qubit. We do not rely on the rotating-wave or dispersive approximations, but solve the full transverse interaction between the qubit and the cavity mode. We show that by carefully choosing the detuning and interaction time, we can exploit a recurrence in the qubit-cavity dynamics in a way that makes it possible to perform very fast, high fidelity, QND measurements. Here, the qubit measurement is performed more like a gate operation between the qubit and the cavity, where the cavity state can be amplified, squeezed, and released in a time-sequenced fashion. In addition, we also show that the non-demolition property of the off-resonant approximation breaks down much faster than its dispersive property,  suggesting that many of the dispersive measurements to date have been implemented outside the QND regime.   
\end{abstract}

\maketitle

% --------- expanding the abstract --------- 
The ability to measure quantum systems in a fast and reliable manner is important in most future quantum technologies, including quantum computing~\cite{schoelkopf2004,schoelkopf2007,fowler12, devo13,schoelkopf2007a}. For superconducting systems, the state of the art qubit measurements involve coupling the qubit to a single cavity mode so as to produce a phase shift of the cavity dependent on the qubit state. This phase shift can then be amplified and measured via homo- or heterodyne detection. It has been standard practice to couple the qubit to the cavity using a static {\it transverse} interaction~\cite{simmonds2014, simmonds2014b, schoelkopf2005, wallraff2017, hanggi2009, clerk2017, blais2009, martinis2014} which, while relatively simple to engineer, will only provide a quantum non-demolition (QND) measurement~\cite{caves1980} if the qubit-cavity detuning is much larger than the coupling rate. QND measurements provide information about the qubit state without generating any additional disturbance to the qubit, maintaining the fidelity of the measurement result~\cite{wallraff2017}, key when making continuous, weak qubit measurements. Achieving a QND measurement by employing a large detuning reduces the speed of the measurement~\cite{blais2017,krantz2016}. A new QND technique has recently been suggested to overcome this limitation by engineering a new qubit-cavity circuit with a {\it longitudinal} interaction that is actuated by a parametric drive~\cite{blais2015}. Here we show that by controlling in time the dynamics of the full transverse interaction between a qubit and a cavity mode, it is possible to make fast, high-fidelity qubit measurements that satisfy the QND criterion. Implementing this measurement requires the ability to smoothly turn on and off the interaction strength between the qubit and the cavity~\cite{simmonds2014, simmonds2014b}, so that a recurrence phenomenon in the dynamics continuously displaces the cavity state, while periodically returning the qubit to its original state. The measurement circuit we construct can also enhance the fidelity and speed of the measurement through the in-situ amplification and generation of a squeezed state in the cavity. We will consider the measurement of a multi-level ``transmon'' qubit~\cite{Koch07, Houck2009}, as well as an ``ideal'' qubit with exactly two energy levels. We find that, for a given coupling strength, the measurement of the former takes approximately twice as long as that for the latter. 

A transverse interaction between a superconducting qubit and a microwave cavity mode is described by the Hamiltonian 
\begin{equation}
    H = \hbar \omega_{\ms{c}} a^\dagger a + \hbar \frac{\omega_{\ms{q}}}{2} A + \hbar g B (a+a^\dagger), 
\end{equation}
where $a^\dagger$ is the creation operator of the cavity mode, $\omega_{\ms{c}}$ and $\omega_{\ms{q}}$ are the frequencies of the  cavity and qubit, respectively, and $g$ is the coupling rate~\cite{simmonds2010, simmonds2014, simmonds2014b, korotkov2013}. For an ideal qubit $A = \sigma_z$ and $B = \sigma_x = \sigma_+ + \sigma_-$, where $\sigma_z$ and $\sigma_x$ are the Pauli $z$ and $x$ operators for the qubit. For a transmon $A =2 b^\dagger b - \delta b^\dagger b(b^\dagger b-1)/\omega_q$ with anharmonicity $\delta>0$ and $B = b+b^\dagger$ in which $b$ is the annihilation operator~\cite{Koch07, Houck2009}. We will define the states $|0\rangle$ and $|1\rangle$ as the lower and upper eigenstates of the qubit, respectively. 

% ----------------- Dispersive approximation ---------------- 
Before proceeding it is instructive to examine the behavior of the dispersive and rotating-wave approximations for the ideal qubit coupled to a cavity. If we move into the interaction picture with respect to the Hamiltonians of the qubit and a cavity, the transverse Hamiltonian becomes 
\begin{equation} 
H^{(I)}= \hbar g \left[ a \sigma_+ e^{i \Delta t} + a^\dagger \sigma_+ e^{i(\omega_{\ms{c}} + \omega_{\ms{q}}) t} \right] + \mbox{H.c.} 
\label{eq:rabiham}
\end{equation}
in which $\Delta \equiv \omega_{\ms{q}} - \omega_{c}$ is the detuning between the qubit and cavity mode. The rotating-wave approximation (RWA) applies when $\omega_{\ms{c}}+\omega_{\ms{q}} \gg g$ and $\omega_{\ms{c}}+\omega_{\ms{q}} \gg \Delta$. It eliminates the highest frequency terms, reducing the interaction Hamiltonian to 
\begin{equation}
H^{(I)}_{\ms{RWA}} \approx \hbar g \left[ a \sigma_+ e^{i \Delta t} + a^\dagger \sigma_- e^{-i \Delta t} \right]  
\label{eq:rabiham2}
\end{equation} 
The dispersive approximation (DA) assumes further that $\Delta \gg g$ so that the interaction acts like a perturbation on the Hamiltonian $H_0 = \hbar \omega_r a^\dagger a + \hbar \omega_{\ms{q}} \sigma_z/2$. The perturbation expansion yields
\begin{equation}
  H^{(I)}_{\ms{DA}} \approx \hbar g \left(\frac{g}{\Delta}\right) \sigma_z a^\dagger a + \mathcal{O}\left[ \left(\frac{g}{\Delta}\right)^2 \right]  
  \label{pert}
\end{equation} 
Two useful properties of the dispersive regime are i) that the frequency, and thus the phase, of the cavity depends on $\sigma_z$, providing a convenient way to measure the qubit's state, and ii) that Eq.(\ref{pert}) satisfies the QND condition, $[H^{(I)}_{\ms{DA}}, \sigma_z] = 0$, when $g/\Delta\ll 1$. However, as seen later, there is a much greater restriction on the ratio $g/\Delta$ in order to achieve the QND property than to achieve a sizable state-dependent phase shift. Our numerical simulations reveal that $g \lesssim \Delta/10$ is sufficient to generate a state-dependent phase shift that can discriminate the qubit's state. However, as seen later in Fig.~\ref{fig:szosc1}, the transverse interaction produces oscillations in the state of the qubit whose amplitude scales as $g/\Delta$. With the cavity mode prepared in a coherent state with amplitude $\alpha = 3$ (so that $\langle a^\dagger a \rangle = 9$), preserving the qubit state to within $1\%$ requires $\Delta=30g=3~\mbox{GHz}$, or $g=\Delta/30$. We believe this point has not been emphasized in previous work. (Further details regarding the breakdown of the rotating-wave approximation are given in the supplemental material~\cite{supp}). 

The dispersive regime also has some disadvantages. First, the condition $g \ll \Delta$ limits the speed of the measurement, both because it limits the size of $g$, even for the largest values of $\Delta$, and because the effective coupling rate in this regime is $g^2/\Delta \ll g$. Another disadvantage comes from the fact that the perturbative expansion in Eq.(\ref{pert}) involves powers of $a^\dagger a$ resulting in a breakdown in the QND condition for large photon numbers in the cavity. This then restricts the number of photons used during, and the resulting signal-to-noise of, a qubit measurement. Given current experimental values of $g$ and $\Delta$ this limit is $\langle a^\dagger a \rangle \lesssim 10$. Finally, for static coupling $g$, the qubit is susceptible to decoherence, from both energy relaxation through the cavity at a rate $(g/\Delta)^2\kappa$ and dephasing from photon shot noise at a rate $(g^2/\Delta)\delta n$. Here, $\kappa$ is the total decay rate of the cavity and $\delta n$ is photon number dependent shot noise in the cavity. 

\begin{figure}[!tb] 
{\includegraphics[width=1\hsize]{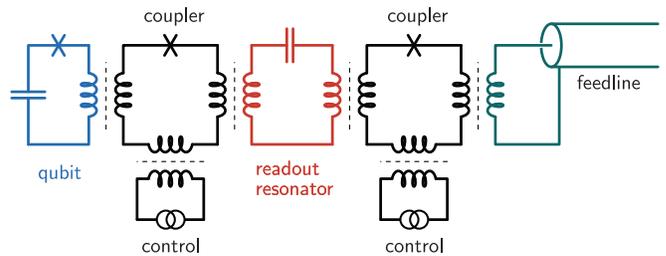}}
\caption{(Color online) The circuit consists of a qubit, a cavity, and a feedline for transmitting the readout signal to room temperature. There are two tunable rf-SQUID couplers, one between the qubit and the cavity and one between the cavity and the feedline. A magnetic flux applied to either coupler can be used to turn ``on" and ``off" the interactions in a controlled manner.
}
\label{fig:circuit1} 
\end{figure} 

In order to overcome all these limitations, we propose the use of tunable couplers to isolate the interactions between the qubit, the cavity, and the measurement feedline~\cite{simmonds2010,simmonds2014,Naik17,Yin2013,korotkov2013,Chen2014}. The full circuit is shown in Fig.~\ref{fig:circuit1}. Each coupler provides a tunable interaction strength through the application of a well controlled magnetic flux through its central loop. This can be achieved both step-wise or through parametric modulation~\cite{simmonds2010,simmonds2014}. In addition, the couplers can allow us to adjust the effective detuning between the qubit and the cavity, as well as to amplify and squeeze the state of the cavity by frequency modulation at $2\omega_{\ms{c}}$, as discussed later. It is easy to see that when $g=0$ for both couplers, decoherence of the qubit from the cavity is eliminated and the decay of the cavity state is dominated by its intrinsic loss rate, $\kappa\rightarrow\kappa_i$. What's less obvious is that fast QND measurements can be achieved with this circuit even when using large photon numbers in the cavity. This somewhat surprising result requires going outside of the dispersive approximation so as to use the full transverse interaction. This also makes our qubit readout process, as discussed below, more unconventional when compared to typical dispersive measurements. 
	
In standard dispersive readout, the cavity's decay rate, $\kappa$, plays an important role in discriminating the qubit's state~\cite{wallraff2017}. It determines the timescale over which the qubit information is retrieved, the resulting signal size with respect to the dispersive shift, and it has to be balanced against increased qubit loss through the Purcell effect~\cite{Houck2008}. For our scheme, we can essentially ignore the cavity decay rate and assume $\kappa=\kappa_i\simeq 0$ when analyzing qubit state discrimination; $\kappa$ is only made large when the state information inside the cavity needs to be retrieved, at which point the qubit is decoupled from the cavity and is immune to cavity induced decoherence. Further analysis of the affects of cavity losses can be found in the supplemental material~\cite{supp}.

\begin{figure*}
{\includegraphics[width=1\hsize]{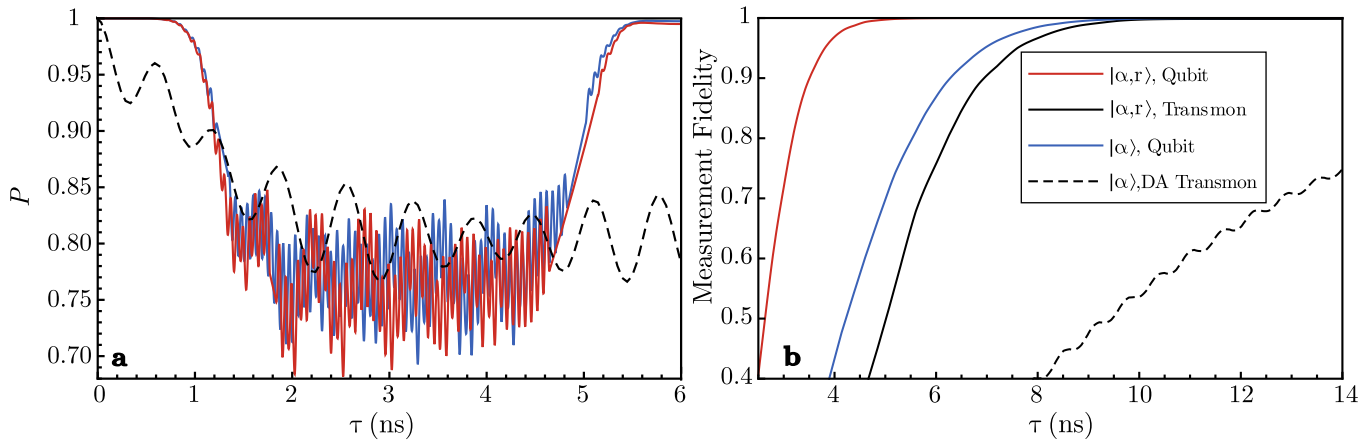}}
\caption{(Color online) a) The excited-state population of the qubit during the interaction with the cavity, when the qubit is initially excited. Here the two optimized protocols use a maximum interaction strength $g/2\pi = 100~\mbox{MHz}$ and an optimal choice of the cavity and qubit frequencies, $\omega_c=8.128~\mbox{GHz}$ and $\omega_q=6.998~\mbox{GHz}$. The dashed curve represents a typical, standard dispersive measurement of a transmon~\cite{wallraff2017}, in which the cavity is quickly driven to a coherent state of amplitude $|\langle a \rangle| = \sqrt{2.5}$ and the always-on coupling strength is $g/2\pi=208~\mbox{MHz}$. The light-blue curve is our optimized protocol in which the cavity starts in a coherent state with $|\langle a \rangle|  \approx 8.15$. The red curve is also an optimized protocol but this time using a squeezed coherent state with squeezing parameter $r=1$ and $|\langle a \rangle|  \approx 8.15$. b)  Fidelities for the measurement of a transmon and an ideal qubit, as quantified by a measure of the distinguishability of the two final cavity states (see text), as a function of the interaction time, $\tau$. Red line: ideal qubit in which the cavity state is squeezed with $r = 1$ and $|\langle a \rangle| \approx 8.15$; light-blue line: same but the cavity state is coherent with $|\langle a \rangle| \approx 8.15$; black line: transmon using an optimal protocol and a cavity squeezed state with $r = 1$ and $|\langle a \rangle| \approx 8.15$; dashed line: transmon using a typical protocol within the dispersive approximation (DA) and a coherent state with $|\langle a \rangle| = \sqrt{2.5}$. All solid curves in b) are optimized to give the maximum distinguishability with the disturbance to the qubit limited to $0.5\%$ (see text)}. 
\label{fig:szosc1}
\end{figure*}
Our qubit readout process is pulsed in time much like a logical gate sequence. First, we excite the cavity mode with a resonant tone. The cavity state can also be prepared in a squeezed state by pumping at $2\omega_c$. Next, we smoothly turn on the coupling between the qubit and the cavity for a given evolution time and then turn it off. This interaction correlates the state of the qubit with that of the cavity. Then, we can apply an amplification to the cavity with a tone at $2\omega_c$. Finally, we perform the readout by applying a coupling pulse between the cavity and the feedline, increasing $\kappa$ to a large value, launching the final cavity state on the order of $10$~ns, to be recorded at room temperature after passing through the rest of the amplification chain. The disturbance to the qubit and the ultimate achievable fidelity can be obtained by modeling the full qubit-cavity interaction over its evolution time. The interaction correlates the $\sigma_z$ basis of the qubit with two states of the cavity, such that the latter are separated by a large phase shift. Perfect measurement fidelity occurs, assuming an ideal amplifier chain, when the two possible qubit states are perfectly correlated with two perfectly orthogonal cavity states. To write this formally, to achieve an ideal measurement, the joint state of the qubit-cavity system must be transformed as $|x\rangle |S\rangle \rightarrow | \phi_{x} \rangle |\psi_{x}\rangle$, in which $\langle \psi_0 | \psi_1\rangle = 0$ and $| \phi_{x} \rangle = | x \rangle$, for an initial qubit state $|x\rangle$ (with $x=0,1$ representing the logical basis) and cavity state $|S\rangle$. These two conditions ensure that a measurement of the cavity can infer the initial state of the qubit without error (perfect fidelity) and that the measurement leaves the $\{|0\rangle,|1\rangle\}$ basis of the qubit undisturbed. The ability to evolve this system exactly so that these conditions are met within some tolerance provides us with a new way of achieving a high fidelity, QND measurement. 

In general the evolution generated by the interaction will not achieve a perfect QND measurement. A natural definition of the fidelity of a measurement is $F=1-p_{\ms{e}}$ where $p_{\ms{e}}$ is the probability that the measurement result does not correctly reflect the initial state of the qubit. Given that the initial qubit state $|x\rangle$ generates the final cavity state $\rho_x$, $p_{\ms{e}}$ is limited by the distinguishability of $\rho_0$ and $\rho_1$~\cite{FuchsPhD}. For convenience here we use a slightly different definition of the fidelity in which $p_{\ms{e}}$ is replaced with the closely related quantity $\mathcal{F} \equiv \mbox{Tr}[\sqrt{\sqrt{\rho_1}\rho_0\sqrt{\rho_1}}]$, which is a measure of the  indistinguishability of $\rho_0$ and $\rho_1$~\cite{FuchsPhD, Fuchs99, Jozsa95}. 

We define the disturbance induced by the measurement as $d = \max(p_0,p_1)$ in which $p_x$ is the probability that the qubit initially in state $|x\rangle$ is flipped to $|1-x\rangle$ by the interaction. Since the interaction is unitary, the Schmidt decomposition allows us to write the final joint state as 
\begin{equation} 
|\Psi_x\rangle = \sqrt{1-\epsilon_x} | \phi_{x}^{+} \rangle |\psi^{+}_{x}\rangle + e^{i\theta}\sqrt{\epsilon_x}  | \phi_{x}^{-}\rangle |\psi^{-}_{x}\rangle , 
\end{equation} 
in which $\langle \phi_{x}^+ | \phi_{x}^- \rangle = \langle \psi_{x}^+ | \psi_{x}^- \rangle = 0$ and $\theta$ is an arbitrary phase. If $| \phi_{x}^{+} \rangle$ is the desired final state for the qubit then $\epsilon_x$ is an error probability that contributes both to the infidelity and disturbance. We can further decompose $| \phi_{x}^{+} \rangle = \sqrt{1-q_x}|x\rangle + e^{i\phi}\sqrt{q_x}|1-x\rangle$, and the total probability of the flip $|x\rangle \rightarrow |1-x\rangle$ is then $p_x = \epsilon_x + q_x - 2 \epsilon_x q_x$.

In order to determine the ability of the qubit-cavity transverse interaction to implement a high fidelity QND measurement, we use a numerical search to find the optimal values for $\omega_{\ms{q}}$, $\omega_{\ms{c}}$ (and thus $\Delta$), and the coupler pulse shape for a fixed choice of the interaction time, $\tau$, maximum interaction strength, $g$, and the initial cavity state $|S\rangle$ (see supplemental material~\cite{supp}). These optimal values maximize $F$ while constraining $d$ to remain below a specified bound. We perform the numerical optimization across a range of $\tau$ to determine the time required to reach a given fidelity for a given level of disturbance. For the case of the transmon, we choose a typical anharmonicity of $\delta=200~\mbox{MHz}$. For $|S\rangle$ we explore both coherent and squeezed states~\cite{carmichael2005, Schumaker1986}, optimizing the squeezing angle of the latter. Squeezed states are able to resolve a smaller phase shift, thus reducing the time required to achieve a given fidelity.     

The numerical search reveals two key features of our qubit readout process which hold both for the ideal and the transmon qubit. First, the full transverse interaction possesses a recurrence-like behavior that undoes the effect it has on each of the qubit basis states $|0\rangle$ and $|1\rangle$ at a specified evolution time, thus implementing a QND measurement. Second, as expected for a given interaction time $\tau$, the distinguishability of the final cavity states increases both with the amplitude and the degree of squeezing of the initial cavity state. For large amplitudes ($|\langle a \rangle|\gtrsim 5$), we can achieve fidelities of $F>99\%$ in measurement times of $\tau\sim10$~ns with $g/2\pi = 100~\mbox{MHz}$. In this regime, we find an optimal detuning of $\Delta/g\approx10$. To obtain the same fidelity using lower amplitudes in the cavity (such as those typically used in the dispersive approximation to achieve the QND aspect) would require significantly longer measurement times. Our time-dependent coupling allows us to use larger amplitudes and thus increase the measurement speed without sacrificing its QND aspect.

In Fig.~\ref{fig:szosc1}(a) we plot the controlled evolution of the excited state population $P$ of the ideal qubit, which shows the recurrence-like behavior during the transverse interaction. For the optimized protocol shown, the qubit is initially placed in $|1\rangle$, we choose an interaction time of $\tau = 6~\mbox{ns}$, a maximum interaction strength of $g/2\pi=100$~MHz, and two different initial states of the cavity. The first is a coherent state with $|\langle a\rangle| \approx 8.15$ and the second is equally displaced, but with a squeezing parameter of $r=1$ (see supplemental material~\cite{supp}). For these protocols, we constrain the expectation value of $\sigma_z$ to return to within $1\%$ of its initial value. This corresponds to limiting the probability of a flip to $p_x \leq 0.5\%$. The plot shows the evolution of the qubit state as it migrates away, and then finally returns to its initial state. While both protocols are similarly effective at leaving the qubit undisturbed, the squeezed state achieves a fidelity of $D=99.76\%$ in just $\tau = 6~\mbox{ns}$, whereas the coherent-state protocol requires at least twice as long. For comparison, we also show in Fig.~\ref{fig:szosc1}(a) the evolution for a protocol similar to a standard dispersive readout with a static coupling strength of $g/2\pi=208$~MHz, $\Delta/g >$ 7.5, and initial cavity states with a lower amplitude, $\langle a \rangle = \sqrt{2.5}$. In an ideal case, the measurement is turned on by quickly pumping the cavity up to its maximum amplitude and then letting it decay to an output channel (the presence of the output channel necessarily hinders the measurement rate, therefore our standard model is an ideal consideration). Notice that the standard dispersive readout disturbs the qubit significantly with some recurrence behavior, but never returns the qubit to its initial state.

In Fig.~\ref{fig:szosc1}(b), we plot the fidelities achieved by our three protocols as a function of the interaction time, $\tau$. We see that moderate squeezing significantly reduces the time required to achieve a given fidelity. We also plot the result for a transmon qubit under the same conditions as the ideal qubit with a squeezed initial cavity state. For the transmon, we find that the main effect of the additional levels is merely to increase the evolution time required to reach the same measurement fidelity. This is about twice that of the ideal qubit. This appears to result from the fact that the third level of the transmon, which is populated during the evolution, generates a third state for the cavity that reduces the final distinguishability (see supplemental material~\cite{supp}). 

\begin{figure}[!tb]
{\includegraphics[height=0.7\columnwidth,width=1\hsize]{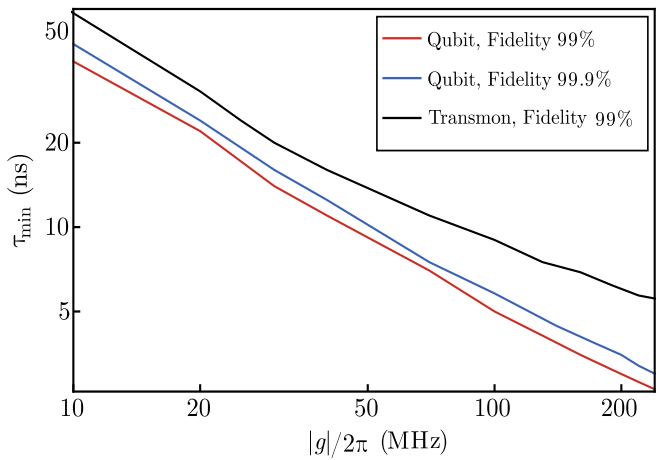}}
\caption{Here we show the minimum time required to achieve 99\% and 99.9\% fidelity as a function of the qubit-cavity coupling rate, $g$, while keeping the probability of disturbing the initial state of the qubit below $0.5\%$. Here, the cavity is initially prepared in a squeezed state. Red: ideal qubit, 99\%; blue: ideal qubit, 99.9\%; black: transmon, 99\%.}
\label{fig:fid1}
\end{figure}

Of particular interest is the time $\tau$ required for the interaction to achieve a given fidelity, while maintaining a minimum disturbance, as a function of the maximum coupling rate between the qubit and the cavity. In Fig.~\ref{fig:fid1}, we plot $\tau$ versus $g/2\pi$ for a qubit disturbance of $d \leq 0.5\%$, both for the ideal qubit and the transmon. As expected, for all cases shown, as the maximum coupling strength is increased, the interaction time can be reduced. For the transmon qubit, interaction times below $10$~ns can be achieved for interaction strengths greater than $100$~MHz. If we consider that further amplification in the cavity and releasing the cavity state takes and additional $20$~ns, then the full readout protocol for a transmon could take $<30$~ns for large interaction strengths. If the output signal from the cavity contains quantum noise only, as expected due to the cold cavity environment, the fidelities we calculate should be experimentally achievable.  

Last but not least, since the protocols we present require choosing specific relationships between certain system parameters, as well as controlling the interaction strength, it is important that these protocols are sufficiently robust to control errors. We have confirmed this and the details are given in the supplemental material~\cite{supp}.

To conclude, we have developed a circuit with a tunable coupling that can utilize a transverse qubit-cavity interaction to make fast, high-fidelity, QND measurements via a pulsed readout protocol that is robust to control errors. In addition, by isolating the qubit and cavity, we can eliminate decoherence of the qubit due to unwanted cavity interactions like the Purcell effect and photon shot-noise dephasing. We find that with a maximum qubit-cavity interaction strength of $g/2\pi = 100~\mbox{MHz}$, it is possible to discriminate a transmon's state with a fidelity of $99\%$ and a QND disturbance no larger than $0.5\%$ in a time of $\sim 10$~ns. For an ideal two-level qubit, this can be achieved in half that time. Because the transmon's relative due purely to leakage to higher levels, it is an interesting question for future work as to whether this effect could be eliminated by using counter-diabatic driving methods~\cite{delCampo13}.   

\textit{Acknowledgments:---}This research was sponsored by the Army Research Laboratory and was accomplished under Cooperative Agreement Number W911NF-16-2-0170.\footnote{The views and conclusions contained in this document are those of the authors and should not be interpreted as representing the official policies, either expressed or implied, of the Army Research Laboratory or the U.S. Government. The U.S. Government is authorized to reproduce and distribute reprints for Government purposes notwithstanding any copyright notation herein.} This research was also sponsored by the National Institute of Standards \& Technology (NIST) Quantum Information Program and was supported by Agreement Number NSA-RA-16-0005.\footnote{Contributions to this work by NIST, an agency of the U.S. Government, are not subject to U.S. copyright.} 

\pagebreak

\begin{center}
\begin{widetext}
\textbf{\large Supplement to  ``Fast, High-Fidelity, Quantum Non-demolition Readout of a Superconducting Qubit Using a Transverse Coupling''}
\end{widetext}
\end{center}

\section{Form of the time-dependent coupling}
\begin{figure}[!htb]
	{\includegraphics[width=1\hsize]{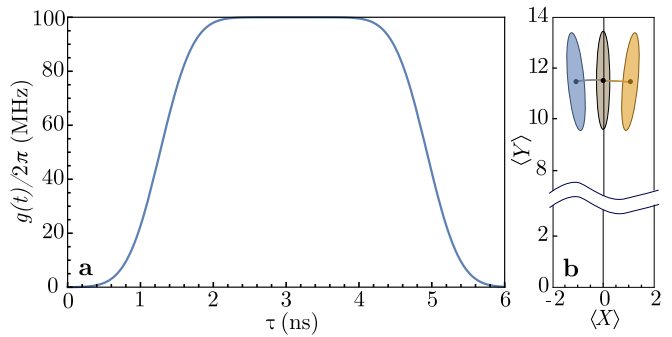}}
	\caption{(Color online) Here we give some further details of the protocol that generates the curves shown in Fig.~2 of the main text: a) the time-envelope of the interaction strength, $g(t)$;  b) a phase-space representation of the initial squeezed state of the cavity, along with the final ($t=6$~ns) states of the cavity resulting, respectively, from the qubit being initially in state $|0\rangle$ and $|1\rangle$. The curves that join the centroid of the initial state to those of the two final states show the path taken by the centroid in the two cases.}
	\label{fig:szosc2} 
\end{figure}

We choose a pulse-shape for the turn-on/turn-off of the transverse interaction (the time-envelope of the interaction strength $g(t)$) that is easy to characterize mathematically but sufficiently flexible for optimization: we want the ability to hold the value of $g$ at its maximum for a given duration, and also to vary the rate at which the interaction is turned on and off. To satisfy these requirements we use the function  
\begin{equation}
g(t)=(g_{\ms{max}}/4)\textrm{Erfc}(-v_1[t-t_1])\textrm{Erfc}(v_1[t-t_2]),
\end{equation}
in which $\textrm{Erfc}(s)=(2/\sqrt{\pi})\int_s^\infty e^{-t^2}dt$. The parameter $v_1$ determines the rate of the turn-on/turn-off, and $t_2-t_1$ (FWHM) characterizes the duration of the interaction. The finite slope of the turn-on/turn-off reduces the sensitivity of the protocol to timing errors (see below). We note that this is one particular choice of the time dependent coupling but any sufficiently smooth function will produce similar behavior (e.g. a piecewise defined function with a Gaussian for the two ramps and a constant value for the mid-region). We show a typical optimal setting for the pulse parameters in Fig~\ref{fig:szosc2} a). This choice of pulse parameters gives rise to the cavity dynamics shown in Fig~\ref{fig:szosc2} b), where the two possible cavity squeezed states are well separated and distinguishable. %Alternatively, one can describe the interaction time as the integration time of the detector, assuming that a noiseless homodyne measurement of the cavity is made. This choice gives us a hybrid of a smooth pulse, with a constant ``top'', which behaves well numerically and does not rely on arbitrarily precise, fine pulse structures. 

\section{Robustness of the protocol}

\begin{figure}[!tb]
	\raggedright
	{\includegraphics[height=0.7\columnwidth,width=1\hsize]{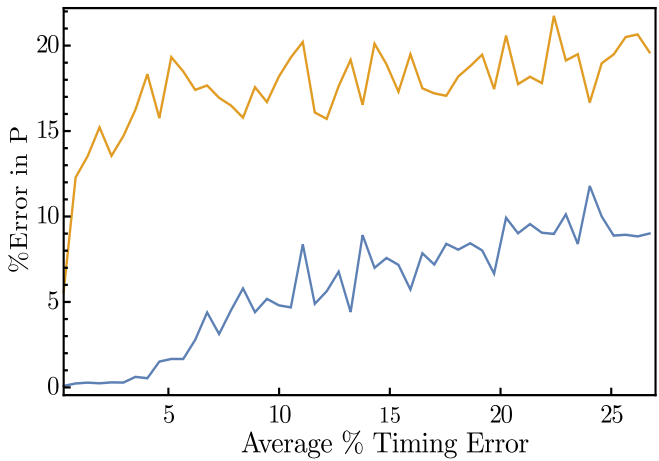}}
	\caption{Here we show the percentage disturbance induced in $P$ by errors in the control parameters for the optimized protocol and fixed parameters $\tau=5$~ns, $\omega_c/2\pi=8.264$~GHz, $\omega_q/2\pi=6.998$~GHz, $g/2\pi=100$~MHz. Blue: smooth envelope for $g(t)$; Orange: square envelope for $g(t)$.}
	\label{fig:error1}
\end{figure}

Since our protocol requires a specific choice for the envelope parameters for a given value of $g$, it is important that the performance is not overly sensitive to fluctuations in these parameters. In Fig.~\ref{fig:error1} we show, for an ideal qubit, the average fractional error induced in the final value of the population of the qubit, $P$, as a function of the average percentage error simultaneously introduced into the three parameters $v_1$, $t_1$, and $t_2$. We also plot this error when the smooth pulse is replaced by a square pulse, showing the necessity of using a function with a smooth ramp. For the case of our smooth pulse, random settings for $v_1$, $t_1$, and $t_2$ are chosen from a Gaussian distribution, centered on previously found optimal values. For a square pulse, we randomly sample from a  Gaussian distribution centered on a sufficiently large $v_1$ to approximate a nearly immediate ramp. 

\section{Numerical Simulation}

Implementing a numerical search requires many simulations of the qubit-cavity system. To reduce the numerical resources we use the equations for a complete set of moments up to third order, replacing the third order moments using the Gaussian approximation. %$\langle  A  B  C \rangle\approx \langle  A  B \rangle\langle  C \rangle+\langle  A  C \rangle\langle  B \rangle+\langle  B  C \rangle\langle  A \rangle-2\langle  A \rangle\langle  B \rangle\langle  C \rangle$. 
%It turns out that 
This results in 14 coupled differential equations both for the ideal qubit and the transmon. We employ these approximate moment equations in our numerical search, and then use an exact simulation to determine the evolution generated by the parameters so obtained.

The constraints we place on the system parameters are chosen to be realistic for the circuit shown in Fig.~1 of the main text. These constraints are $3~\mbox{GHz} \leq \omega_{\ms{q}}/2\pi \leq 7~\mbox{GHz}, 8~\mbox{GHz} \leq \omega_{\ms{c}}/2\pi \leq 11~\mbox{GHz}, g/2\pi \leq100~\mbox{MHz}$, $\gamma/2\pi=10~\mbox{kHz}$ (where $\gamma$ is the decay rate of the cavity). These constraints allow the search to explore a realistic parameter space, without causing numerical search issues due to a overly large search space.

\section{Breakdown of the Dispersive Approximation}
As discussed by Blais~\etal~\cite{blais2009}, care must be taken when employing the dispersive approximation~\cite{korotkov2013,wallraff2017}. Unfortunately, the breakdown of the QND nature of this approximation has not been clearly articulated in recent analyses. Under the dispersive approximation (DA) the state of a cavity mode will rotate in phase space at different speeds depending on whether the qubit is in state $|0\rangle$ or $|1\rangle$. The DA thus predicts that the interaction between a qubit and cavity will generate a certain distance (and thus distinguishability), $\lambda$, between the final states of the cavity. This distance is the key quantity that determines the fidelity of the measurement achieved by the qubit/cavity interaction. Separately, the QND nature of the measurement is determined by the way in which the interaction changes the initial states $|0\rangle$ and $|1\rangle$ of the qubit. For a good QND measurement the interaction should leave the qubit in its initial state for both possible initial states. Under the DA the interaction is assumed to be done weakly, in an attempt to leave the qubit in its initial state, achieving a QND measurement. 

We show here that while the DA requires the condition $\Delta\gg g$, the predictions of the DA regarding the QND nature of the measurement break down at larger $\Delta$ than the prediction of measurement fidelity. Measurement fidelity predictions of the DA agree to within 1\% of our exact simulation, as long as one maintains $\Delta/g\gtrsim 20$, for the example case of a cavity coherent state with $\alpha=3$. However, restricting to the same 1\% agreement, the dispersive approximations prediction of a QND measurement breaks down at $\Delta/g\gtrsim 40$. For the example cases we consider in this section, we utilize a standard approach in the DA, which employs quickly driving the cavity, on resonance, with a simple Gaussian drive pulse and a constant (no time dependence) coupling between the cavity and qubit present. This drive pulse envelope is given by $e^{-(t-t_1)^2/2\sigma^2}$ with amplitude $g_d$.

The dispersive Hamiltonian for the interaction between a qubit and a cavity mode is given by
\begin{equation}
  H^{(I)}_{\ms{DA}} \approx \hbar g \left(\frac{g}{\Delta}\right) \sigma_z a^\dagger a + \mathcal{O}\left[ \left(\frac{g}{\Delta}\right)^2 \right] ,
 \label{pert}
\end{equation} 
where $a$ is the annihilation operator for the cavity mode, $\sigma_z$ is the Pauli z operator for the qubit, and $g(t)$ is a time-dependent coupling rate. This Hamiltonian imparts a phase shift to the cavity mode that depends on the state of the qubit.  It also clearly allows a QND measurement of the qubit as $\sigma_z$ is a constant of motion. The acquired phase shift between the final states of the cavity in this approximation is $\phi=\int_0^\tau dt~g^2/\Delta $ and the resulting distance between the states is $\lambda=2|\alpha| \sin|\phi|$, where $\alpha$ is the amplitude of the cavity state and $\tau$ is the measurement time. These approximations are valid under the conditions $\Delta/g\gg 1$ and $\bar{n}\ll n_{\textrm{crit}} =\Delta^2/4g^2$ where $\bar{n}$ is the average photon number in the cavity mode. 

\begin{figure}[b] 
{\includegraphics[width=1\hsize]{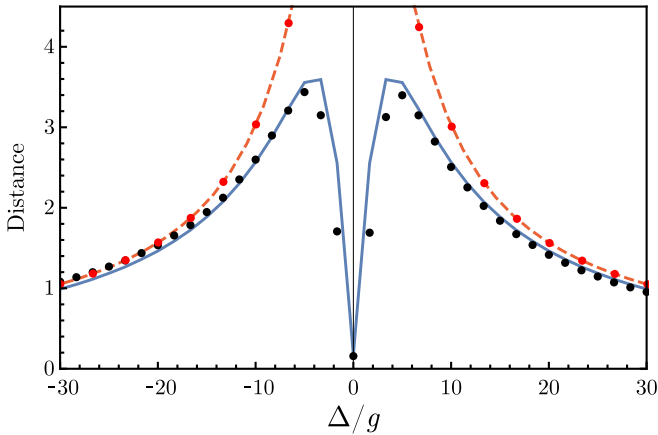}}
\caption{(Color online) The distance between the phase-space centroids of the final cavity states, as a function of $\Delta/g$ with and without the dispersive approximation (DA): Orange dashed curve: DA; Blue solid curve: no DA. The dots give the equivalent results for simulations using our truncated moment equations: Red dots: DA; Black dots: no DA. For these plots we use $g/2\pi=30~\mbox{MHz}$, and the drive envelope is given by $g_d/2\pi=0.604~\mbox{MHz}$, $\sigma=5.424~\mbox{ns}$, $t_1=1.233~\mbox{ns}$, $\tau=32~\mbox{ns}$. These drive parameters drive the cavity to an amplitude of $|\alpha|\approx3$ in $\sim15$~ns.}
\label{distance} 
\end{figure} 
\begin{figure}[!tb] 
{\includegraphics[width=1\hsize]{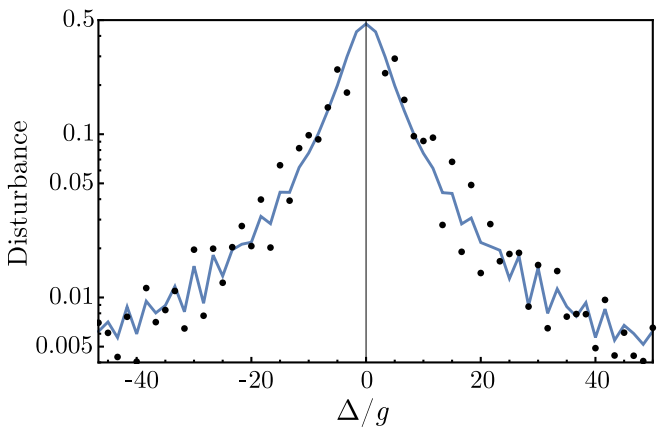}}
\caption{(Color online) The probability that an initial state $|0\rangle$ or $|1\rangle$ flips to the complementary state during the measurement as a function of detuning. The solid line is the exact simulation while the dots show the result of our moment truncations. A dispersive model predicts that the initial states $|0\rangle$ or $|1\rangle$ will remain unchanged during the measurement. These plots are generated using the same parameters as those in Fig.~\ref{distance}.
} 
\label{sz1} 
\end{figure} 

We now compare simulations using the DA (Eq.~(\ref{pert})) to those in which we use full Rabi Hamiltonian so that the effective interaction Hamiltonian is 
\begin{equation}
H^{(I)}= \hbar g \left[ a \sigma_+ e^{i \Delta t} + a^\dagger \sigma_+ e^{i(\omega_{\ms{c}} + \omega_{\ms{q}}) t} \right] + \mbox{H.c.} . 
\label{eq:rabiham}
\end{equation}
In Fig.~\ref{distance} we plot the final distance in phase space, $d$, between the centroids of the cavity mode states with and without the dispersive approximation, as a function of $\Delta/g$. This distance only provides a rough guide to the distinguishability (it ignores the possible improvement from squeezing), but it serves to show how the DA breaks down. For these plots we use $g/2\pi=30~\mbox{MHz}$, drive the cavity to a coherent state with $|\alpha|=3$, and allow the interaction to act for a time $\tau=32~\mbox{ns}$. We see that near $\Delta/g\approx 10$ the predictions of the DA start to deviate significantly from those of the more accurate interaction Hamiltonian. In Fig.~\ref{distance}, for both interaction Hamiltonians we also show the result of an exact simulation (shown by the solid and dashed curves) against the truncated moment equations (shown by the dots) that we use to reduce the numerical overhead.

%*** Here we define the distance of oscillator states to simply be $\sqrt{(\textrm{Re}\langle a\rangle_1-\textrm{Re}\langle a\rangle_0)^2+(\textrm{Im}\langle a\rangle_1-\textrm{Im}\langle a\rangle_0)^2}$, where subscripts denote the initial energy level of the qubit. Dots show that our Gaussian moment approximation agree well with exact simulation, in both the dispersive and non-dispersive models. Admittedly, one can take the dispersive approximation up to third order which refines the approximation somewhat~\cite{blais2009}, but it also breaks down as the assumptions are violated (e.g a third order approximation partially captures the ``plateau'' feature near small detunings). 

We now examine the accuracy of the dispersive approximation in predicting the disturbance to the initial states $|0\rangle$ and $|1\rangle$ of the qubit. The DA predicts no disturbance at all, and in Fig.~\ref{sz1} we show the disturbance predicted by the more accurate interaction Hamiltonian. Specifically we show the probability that the qubit will have flipped to the ground state when it began in the excited state. We see that if we wish to restrict this probability to $1\%$ we must have a ratio $\Delta/g\gtrsim 30$.

%As an example, examining the parameters used by Korotkov~\etal~\cite{korotkov2013}, we find that we come to a conclusion that is quite different to theirs. The dispersive approximation, which they employ, predicts a high fidelity measurement with their parameters, but the more accurate interaction Hamiltonian predicts a measurement fidelity of only $~23\%$. 

%Recall that in the dispersive approximation, $\sigma_z$ is a constant of motion, therefore if we assume that the qubit begins the interaction with $1-P=(1-|\sigma_z|)/2=0$ (it begins in either the ground or excited state), then the dispersive model predicts that it will stay at that value throughout the interaction. However, using Eq.~\ref{eq:rabiham}, Fig.~\ref{sz1} shows a very different behavior, as the dispersive assumptions are violated. In this case, disagreement between the dispersive and non-dispersive models begins near $\Delta/g\approx50$, at much larger values of $\Delta/g$ than in the case of the breakdown for distance calculations. 

The heart of the issue of the dispersive approximation, $\Delta \gg g$, is that in order to achieve a fast qubit measurement, it is advantageous to make $g$ as large as practically possible. Depending on the system (the value of $\Delta$), this may push one out of the regime where the dispersive approximation is valid.

\section{Model of the Transmon} 
A transmon is a non-linear oscillator whose lowest two energy states are used as the qubit. The model of the transmon is an inverted Duffing oscillator for which the Hamiltonian is  
\begin{equation}
  H_\textrm{q}=\omega_{\textrm{q}0} b^\dagger b-\frac{\varepsilon}{2}b^\dagger b(b^\dagger b-1),
\end{equation}  
where $\varepsilon\equiv\omega_{\textrm{q}0}-\omega_{\textrm{q}1}>0$ is the ``anharmonicity" and $\omega_{\textrm{q}i}$ is the transition energy between levels $i\rightarrow i+1$. We find that using this model for the qubit lengthens the measurement time required to obtain a given fidelity, but does not significantly impact the QND nature of the measurement (higher levels are populated during the interaction with the cavity mode, but the same recurrence phenomena returns the transmon to its initial state).   

\begin{figure}[!htb] 
{\includegraphics[width=1\hsize]{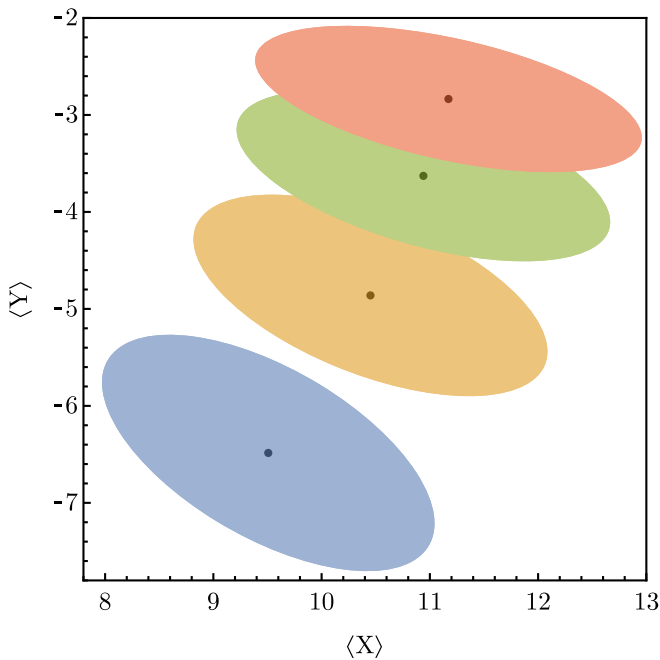}} 
\caption{(Color online) Contour depiction of the phase-space distributions for the final states of the cavity mode generated by the first few levels of the transmon. The initial state of the cavity mode is a squeezed state with $\alpha\approx 8.15$ and $r=1$. Blue: ground state; orange: first excited state; green: second excited state; red: third excited state. The various physical parameters are $g/2\pi=100~\mbox{MHz}$, $\Delta/2\pi=353.71~\mbox{MHz}$, $\varepsilon/2\pi=200~\mbox{MHz}$, and those that determine the shape of the interaction envelope $g(t)$ are  $v_1=1.214~\mbox{ns}^{-1}$, $t_1=2.249~\mbox{ns}$, $t_2=9.551~\mbox{ns}$.
}
\label{phasespace1}  
\end{figure}

\begin{figure}[!htb] 
{\includegraphics[width=1\hsize]{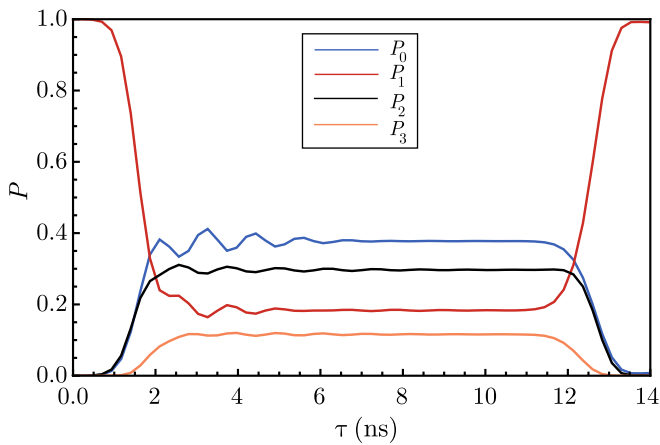}} 
\caption{(Color online) Population of the first four levels of the transmon, during its time dependent coupling interaction with the cavity, where the transmon begins the interaction in the first excited state (red, $P_1$). Significant population is leaked into the ground (blue, $P_0$), second excited state (black, $P_2$) and even the third excited state (orange, $P_3$). Additional levels are included in simulations but maintain minimal ($<3\%$) populations (not shown). System parameters are the same as listed in Fig.~\ref{phasespace1}.
}
\label{tpop1}  
\end{figure}
As compared to the ideal qubit, phase shifts induced in the cavity mode by the various energy levels of the transmon accumulate more slowly. For example, we show in Fig.~\ref{phasespace1} that a 99\% distinguishability between the first two levels of the non-linear oscillator can be achieved in a minimum time of $\tau=14~\mbox{ns}$. For the same parameters, an ideal qubit can achieve the same distinguishability in a minimum time of only $\tau=6~\mbox{ns}$. We also show two of the next higher excited states in the non-linear oscillator, which are partially distinguishable from their nearest levels. 

In the case of the ideal qubit, increasing the photon number in the cavity leads to a strictly increasing measurement fidelity, for a given measurement time. For the transmon model, in addition to optimizing circuit parameters, one also needs to optimize the photon number in the cavity (including squeezing strength and angle $r,\theta$). Once the optimal cavity photon number is found, a high fidelity, QND measurement of the qubit is possible. Fig.~\ref{tpop1} shows that if the transmon begins the interaction in its first excited state, significant leakage to other states occurs. This leakage is a direct result of using a full transmon model, along with large photon number in the cavity. However, when a time dependent coupling is used, we take advantage of a similar recurrence that is present in the ideal qubit model. This recurrence returns the populations to values which are nearly their initial values, constituting a QND measurement.

\vspace{3mm}
\section{Disturbance introduced by the measurement}

Here we give some further explanation about how the disturbance induced by the measurement, being the measure of the ``QND-ness" of the measurement, is related to an explicit form for the final joint state of the qubit and the cavity. Given that the evolution is unitary, the most general form for this final joint state is   
\begin{equation} 
|\Psi_x\rangle = \sqrt{1-\epsilon_x} | \phi_{x}^{+} \rangle |\psi^{+}_{x}\rangle + e^{i\theta}\sqrt{\epsilon_x}  | \phi_{x}^{-}\rangle |\psi^{-}_{x}\rangle , 
\end{equation} 
in which $x=0,1$ denotes the initial state of the qubit in the $\sigma_z$ basis,  $|\phi_{x}^\pm\rangle$ are orthogonal states of the qubit, $|\psi^\pm_{x}\rangle$ are orthogonal states of the cavity mode, and $\theta$ is an arbitrary phase factor. 

Now recall that we would ideally like the qubit to be left in a state $|x\rangle |\chi_{x}\rangle$ where $|\chi_{x}\rangle$ and $|\chi_{1-x}\rangle$ are orthogonal. So if we identify $| \phi_{x}^{+} \rangle$ as the final state that the qubit would be in if the measurement were ideal, and expand the states $| \phi_{x}^\pm \rangle$ as 
\begin{align}
   | \phi_{x}^\pm \rangle = \sqrt{1-q_x}|x\rangle \pm e^{i\phi}\sqrt{q_x}|1-x\rangle , 
\end{align}
then the parameters $\epsilon_x$ and $q_x$ quantify deviations from an ideal measurement. That is, when $\epsilon_x = q_x = 0$ the final joint state is 
\begin{equation} 
|\Psi_x\rangle =  | x \rangle |\psi^{+}_{x}\rangle  
\end{equation} 
so that the measurement will be ideal if it is also true that $\langle \psi^{+}_{x}|\psi^{+}_{1-x}\rangle = 0$. 

Both $\epsilon_x$ and $q_x$ contribute to the probability that the state of the qubit gets flipped from the initial state $|x\rangle$ to the other state $|1-x\rangle$ during the measurement interaction. To obtain the final state of the qubit we trace over the cavity mode, and obtain 
\begin{align}
   \sigma_{\ms{f}} & = (1-\epsilon_x)| \phi_{x}^{+} \rangle \langle \phi_{x}^{+}|  + \epsilon_x| \phi_{x}^{+} \rangle \langle \phi_{x}^{+}| \\
                   & = (1 - \epsilon_x - q_x + 2 \epsilon_x q_x) |x\rangle \langle x| + \ldots 
\end{align}
From this we see that the probability that the qubit is left in state $|1-x\rangle$ at the end of the interaction instead of the initial state $|x\rangle$ is 
\begin{align}
   p_x = \epsilon_x + q_x - 2 \epsilon_x q_x . 
\end{align}
 
Taking the partial trace over the qubit, the final state of the cavity is  
\begin{align}
   \rho_x = (1-\epsilon_x) |\psi_x^{+}\rangle\langle\psi_x^{+}| + \epsilon_x |\psi_x^{-}\rangle\langle\psi_x^{-}| .  
\end{align}
The states $\rho_0$ and $\rho_1$ then determine the minimum possible error probability $p_{\ms{e}}$. That is, if there is no noise introduced by the amplifier chain, then the fidelity of the measurement is limited only by how well $\rho_0$ and $\rho_1$ can be distinguished as set by quantum mechanics. It is the job of the interaction to make these states as orthogonal as possible, and that of the amplification process to make the difference in the respective signals generated by these two states easily recordable by classical digital circuits.  
 
\vspace{3mm}
\section{Cavity Loss}

The cavity encounters photon loss through two main processes, loss through unwanted coupling to the environment and loss through coupling to an output line. For both of these loss cases, we can modify the Heisenberg equations of motion by adding a Linblad term given by $\mathcal{D}[\hat{O}]=\kappa_j/2(2\hat{a}^\dagger \hat{O} \hat{a}-\hat{a}^\dagger \hat{a} \hat{O} -\hat{O} \hat{a}^\dagger \hat{a})$, for any operator $\hat{O}$ and a loss rate $\kappa_j$. In a typical dispersive readout setup, the cavity is strongly coupled to an output line used in the readout process. Due to this strong coupling, the Purcell effect significantly lowers the lifetime of the qubit due to leakage through the output line. To combat this issue, a Purcell filter (an additional cavity) is used in order to limit the leakage to the output line at the qubit frequency. However, the Purcell filter is weakly coupled to the primary cavity and therefore leads to a slower measurement~\cite{korotkov2015}. This also does not address the direct effect that the loss rate has on the measurement fidelity. In the case of the time dependent coupling we consider, the Purcell effect is not present, since the output line is not coupled during the interaction between the qubit and cavity.

\begin{figure}[!tb] 
{\includegraphics[width=1\hsize]{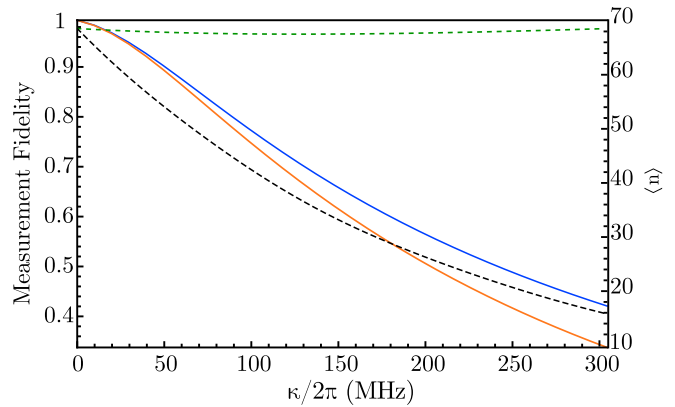}} 
\caption{(Color online)  Measurement fidelity (left axis, solid lines) and photon number in the cavity (right axis, dashed lines), with a sustain pulse (solid blue and dashed green) and without a sustain pulse (solid orange and dashed black). We use an optimized set of parameters for an ideal qubit, and a measurement time of $\tau=5~\mbox{ns}$.
}
\label{loss1}  
\end{figure}

Another reasonable technique is to utilize a sustained drive pulse, whose purpose is to counter the photon loss rate~\cite{martinis2014}. As shown in Fig~\ref{loss1}, a sustain drive can counteract the photon loss, however, it only slightly counters the detrimental effects on the resulting measurement fidelity. In a time dependent coupled system, the internal loss rate of the cavity (without coupling to an output line) can be $\kappa_{int}\approx 10-100~\mbox{kHz}$, while a time independent coupled system is necessarily always coupled to an output line. This output line loss, or external loss, is typical much larger than the internal loss, $\kappa_{ext}\approx 10-50~\mbox{MHz}$. Therefore, with a goal of a fast qubit readout, a small $\kappa$ is desirable.

\bibliography{bib_gjas}

%
% Bibliography
%

\end{document}